\documentclass[a4paper,11pt]{article}
\usepackage{pos}
\usepackage{wrapfig}
\usepackage{enumitem}
\usepackage{pgf}
\usepackage{siunitx}
\usepackage{overpic}

\newcommand{\wmax}{w_{\mathrm{max}}}

\title{The particle-shower simulation code CORSIKA 8}
\ShortTitle{The particle-shower simulation code CORSIKA 8}

\author*[a,b]{Tim Huege}
\author[c]{Maximilian Reininghaus}
\onbehalf{for the CORSIKA~8 Collaboration}

\affiliation[a]{Institute for Astroparticle Physics (IAP), Karlsruhe Institute of Technology, P.O. Box 3640, 76021 Karlsruhe, Germany}
\affiliation[b]{Astrophysical Institute, Vrije Universiteit Brussel, Pleinlaan 2, 1050 Brussel, Belgium}
\affiliation[c]{Institute of Experimental Particle Physics (ETP), Karlsruhe Institute of Technology, P.O. Box 3640, 76021 Karlsruhe, Germany}

\emailAdd{tim.huege@kit.edu}

\abstract{CORSIKA up to version 7 has been the most-used Monte Carlo code for simulating extensive air showers for more than 20~years. Due to its monolithic, Fortran-based software design and hand-optimized code, however, it has become difficult to maintain, adapt to new computing paradigms and extend for more complex simulation needs. These limitations led to the CORSIKA~8 project, which constitutes a complete rewrite of the CORSIKA~7 core functionality in a modern, modular C++ framework. CORSIKA~8 has now reached a state that we consider ``physics-complete'' and a stability that already allows experts to engage in development for specific applications. It already supports the treatment of hadronic interactions with Sibyll~2.3d, QGSJet-II.04, EPOS-LHC and Pythia~8.3 and the treatment of the electromagnetic cascade with PROPOSAL~7.6.2. Particular highlights are the support for multiple interaction media, including cross-media particle showers, and an advanced calculation of the radio emission from particle showers. In this contribution, we discuss the design principles of CORSIKA~8, give an overview of the functionality implemented to date, the validation of its simulation results, and the plans for its further development.}

\ConferenceLogo{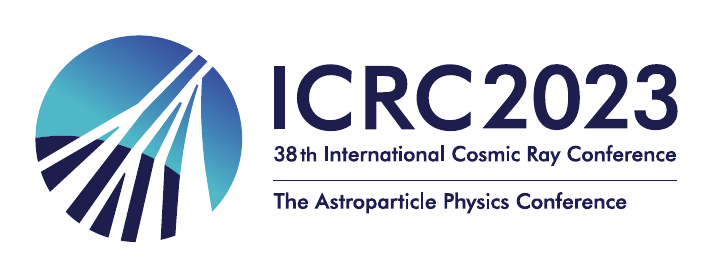}

\FullConference{%
38th International Cosmic Ray Conference (ICRC2023)\\
26 July -- 3 August, 2023\\
Nagoya, Japan}

\begin{document}
\maketitle

\section{Introduction}

The astroparticle physics community heavily relies on Monte Carlo simulations of particle showers in air and other media. For air shower simulations, the de-facto standard for more than 20 years has been the CORSIKA code \cite{HeckKnappCapdevielle1998}, originally developed for the KASCADE experiment. CORSIKA, in its current version 7.75, is still being maintained by KIT; however, its monolithic Fortran structure and the retirement of key developers makes this increasingly difficult. Furthermore, upcoming experiments require more flexibility than the hand-optimized CORSIKA~7 code can provide. In 2018, we therefore started to develop CORSIKA~8, a full rewrite of the CORSIKA core functionality in a C++-based framework \cite{Engel:2018akg}. Since the ICRC2021~\cite{CORSIKA:2021czu}, we have made very significant progress with CORSIKA~8. In the following, we report the design philosophy and current state of the project, showcase some results derived with the code, and give an outlook on the next steps.

\section{Design philosophy and recent progress}

\begin{figure}
    \centering
    \includegraphics[width=0.9\textwidth]{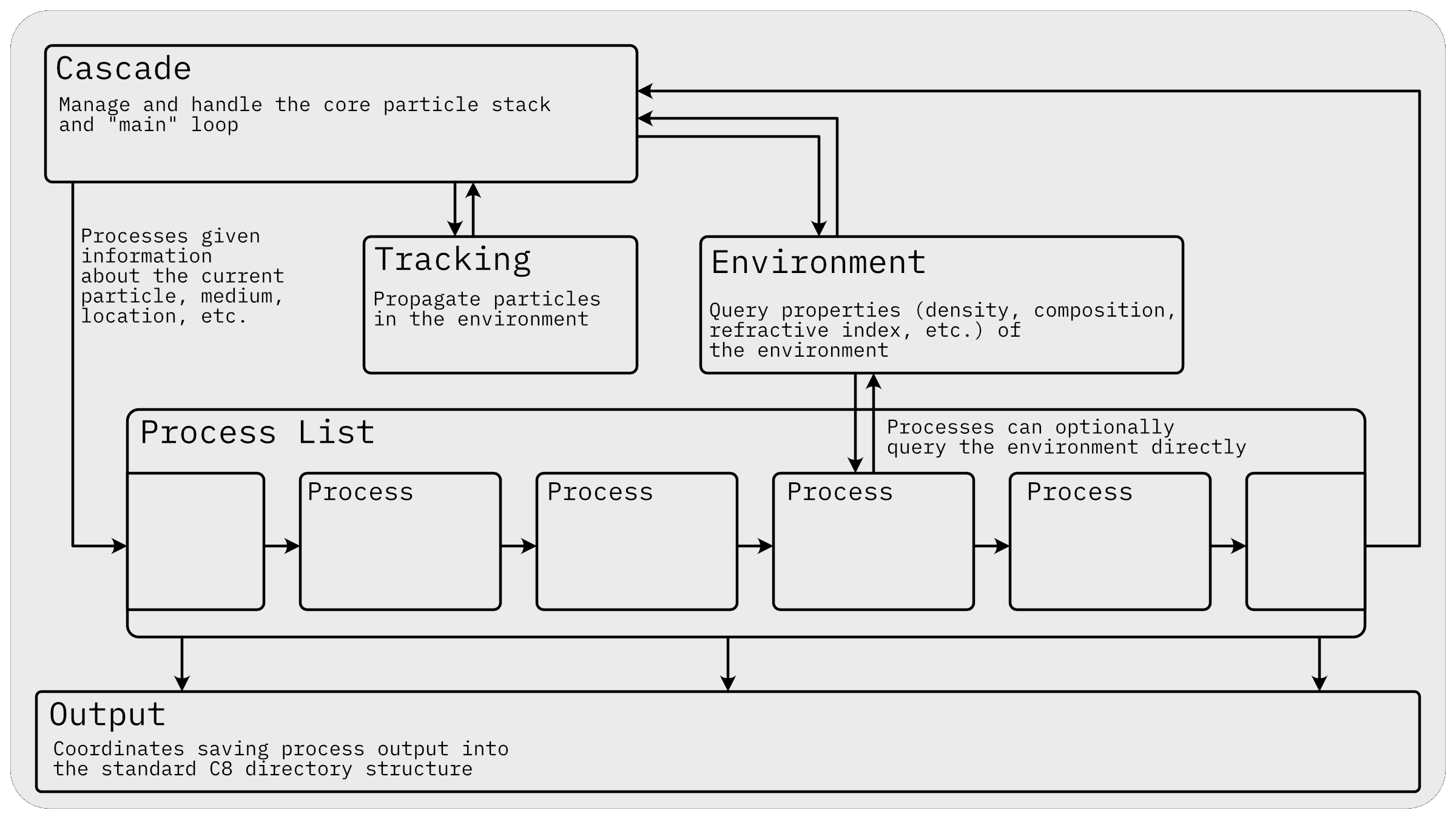}
    \caption{General structure of the CORSIKA~8 code. Please see text for explanation.}
    \label{fig:flowchart}
\end{figure}

CORSIKA~8 is designed as an open-source\footnote{\url{https://gitlab.iap.kit.edu/AirShowerPhysics/corsika/}} community project. KIT is committed to coordinating its development and maintaining key functionality, but the code has been structured as a flexible framework to which individual developers can contribute their functionality in a modular fashion. A general overview of the structure is shown in Fig.\ \ref{fig:flowchart}. The \emph{Cascade} handles the particle stack and main loop. \emph{Tracking} functionality is used to propagate the particles in the environment, taking into account, for example, the deflection of charged particles in magnetic fields. The \emph{Environment} can be queried for all the relevant characteristics of the media that particles propagate through. A major difference with respect to CORSIKA~7 is the flexibility in setting up the environment from geometrical objects (currently spheres and cuboids), with each of them having their own particular media properties. This allows simulations of particle showers in very complex scenarios. For example, we can already simulate showers crossing from air into ice, as detailed in~\cite{JuanICRC2023}. No other simulation code currently offers such flexibility, needed for example by projects aiming to measure radio emission from particle showers in ice.

At the heart of the modular design is the \emph{Process List} which can be assembled flexibly and hosts all the relevant processes such as hadronic interactions, electromagnetic interactions~\cite{AlexanderICRC2023}, decays, checking for transition across media boundaries, radio-emission calculation~\cite{NikosICRC2023}, Cherenkov light calculation (optionally using GPU acceleration)~\cite{DominikICRC2023}, visualization functionality, and more.

Finally, all output of simulation results is handled by a central \emph{Output} manager. Our current data format is based on YAML files for metadata and Apache Parquet files for (larger) binary data. While we might revisit this choice, the underlying data format is not critical, as we provide a Python-based library to transparently access the data from the user side. Simulation steering is envisaged through command line options as well as input files that can be interchanged transparently.

Since the ICRC2021, we have made very significant progress with CORSIKA~8. Among other improvements, we point out the inclusion of FLUKA~\cite{Ferrari:2005zk} as hadronic interaction model at low energies, the inclusion of photohadronic interactions in electromagnetic cascades handled through SOPHIA~\cite{Mucke:1999yb} and Sibyll~\cite{Engel:2019dsg}, the treatment of the LPM effect in electromagnetic cascades, improvements in the handling of multiple scattering in electromagnetic cascades, implementation of particle thinning in electromagnetic cascades with a newly designed algorithm, improved radio-emission calculations in media with realistic refractivity gradients, a proof-of-principle multi-threaded CPU parallelisation of the radio-emission functionality~\cite{AugustoICRC2023} and a preliminary inclusion of Pythia~ 8.3~\cite{Sjostrand:2021dal} as hadronic interaction model at high energies. In the following, we will showcase some current results of CORSIKA~8 in comparison in particular with CORSIKA~7.

\section{Electromagnetic cascades}

Electromagnetic cascades in CORSIKA~8 are being handled by the PROPOSAL \cite{Alameddine:2020zyd} code, currently in version 7.6.2. Details about the simulation of the electromagnetic cascade within CORSIKA~8 and how the implemented physics as well as simulation results compare to EGS4 as used in CORSIKA~7 are provided in reference~\cite{AlexanderICRC2023}. Longitudinal profiles of the electromagnetic component of electron-induced air showers are in agreement on a 10\% level between the two codes. Muons and hadrons produced by photohadronic interactions are approximately 15\% more numerous in CORSIKA~8 than CORSIKA~7, an interesting finding that we will investigate further. (Note that this contribution is subdominant in hadronic air showers.) Energy and lateral distributions are in agreement typically on a 5\% level, with some more pronounced differences at very low energies and very small lateral distances. We point out that while we use CORSIKA~7 as a reference to compare to, some of the implemented physics is different and improved in CORSIKA~8; for example we include triplet production, a process that EGS4 within CORSIKA~7 does not handle. Hence, a 1:1 agreement with CORSIKA~7 is neither expected nor intended. We also note that during our validation efforts, we found and fixed (minor) problems in earlier versions of CORSIKA~7.

Since the ICRC2021, we have implemented particle thinning~\cite{Hillas:1981} for the electromagnetic cascade in CORSIKA~8. Thinning is both most effective in terms of saving computing time and easiest to implement for electromagnetic interactions because of their 1:2 splitting nature. Our implementation is improved with respect to the one used in CORSIKA~7. When a particle has an energy below the \emph{thinning threshold}, $E_\mathrm{th}$, secondaries arising from its interactions are subject to thinning. As long as weight limitation does not set in (see below), one of the two secondaries is randomly chosen to be retained while the other one is discarded. The selection probability $p_i$ of each secondary is proportional to the fraction of its energy with respect to the incoming particle. The weight of the retained particle is increased by a factor $f_i = 1/p_i$ over the weight $w_0$ of the incoming particle. Weight limitation is considered as follows: If at some point the (potential) new weight $w_i$ of a secondary would exceed the user-defined maximum weight, $w_\mathrm{max}$, we resort to \emph{statistical thinning}~\cite{Kobal:2001jx} in which each secondary is considered for retention or removal on its own. In this setting, we have more freedom to alter the retention probabilities as desired. Here, we set
\begin{equation}
    p_i = \max\left(\frac{E_i}{E_1+E_2}, \frac{w_0}{w_\mathrm{max}}\right),
\end{equation}
so that $w_i = {w_0}/{p_i} \leq w_\mathrm{max}$.
As soon as the maximum weight is reached in a particular branch, all particles descending from that vertex are tracked again, having the same weight $w_\mathrm{max}$.

\begin{wrapfigure}{r}{0.61\textwidth}
    \centering
    \begin{Overpic}{\input{weight_distribution.pgf}}\put(57,58){\includegraphics[width=2.2cm]{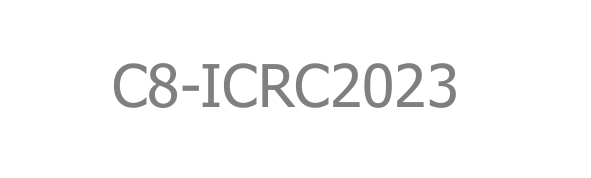}}\end{Overpic}
    \caption{Weight distribution of photons on ground for thinning of the electromagnetic cascade compared between CORSIKA~7 and CORSIKA~8. The narrow weight peaks in CORSIKA~8 minimize artificial fluctuations.}
    \label{fig:thinningweights}
\end{wrapfigure}

Figure \ref{fig:thinningweights} shows a comparison of the obtained weight distributions of photons in  $10^{16}$\,eV photon showers between CORSIKA~8 and CORSIKA~7 using a thinning threshold, $\varepsilon=10^{-5}$, of the primary energy and several maximum weight factors. In the high-weight range, CORSIKA~8 features a narrow peak, while CORSIKA~7 features a broad peak at a value of $w_\mathrm{max} / 2$. The difference is explained by the different implementations of weight limitations. The algorithm chosen in CORSIKA~8 minimizes the artificial fluctuations introduced by the thinning procedure due to narrower weight distributions~\cite{Hansen:2010uk}.

\section{Hadronic cascades}

We offer usage of a wide range of state-of-the-art hadronic interaction models within CORSIKA~8. At high energies, in addition to QGSJet-II.04~\cite{PhysRevD.83.014018}, Sibyll~2.3d~\cite{Engel:2019dsg} and EPOS-LHC~\cite{Pierog:2013ria}, a preliminary inclusion of Pythia~8.3~\cite{Sjostrand:2021dal} is available for testing but still undergoing improvements~\cite{Reininghaus:2023ctx}. At low energies,
the recently included FLUKA~\cite{Ferrari:2005zk} provides increased flexibility in the choice of interaction media and sophisticated modelling in addition to low runtimes. Decays can be handled by Sibyll~2.3d and Pythia~8.3.

\begin{figure}
    \centering
    \begin{overpic}{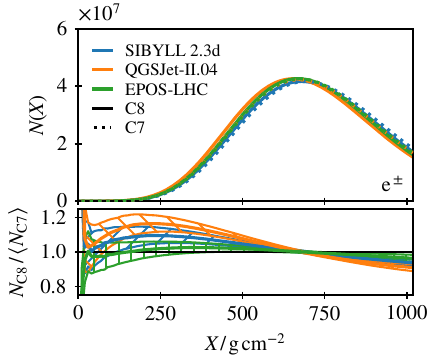}\put(71,74){\includegraphics[width=2.2cm]{watermark.pdf}}\end{overpic}
    \begin{overpic}{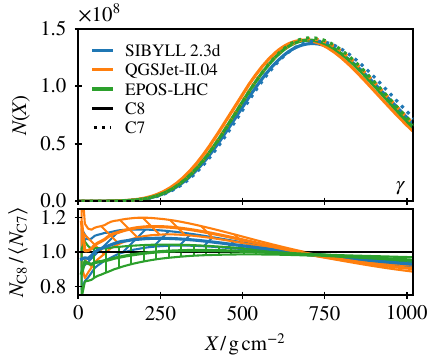}\put(71,74){\includegraphics[width=2.2cm]{watermark.pdf}}\end{overpic}
    \begin{overpic}{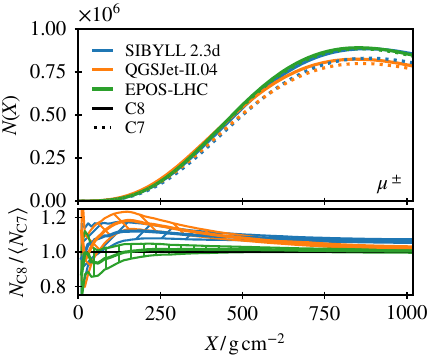}\put(71,74){\includegraphics[width=2.2cm]{watermark.pdf}}\end{overpic}
    \begin{overpic}{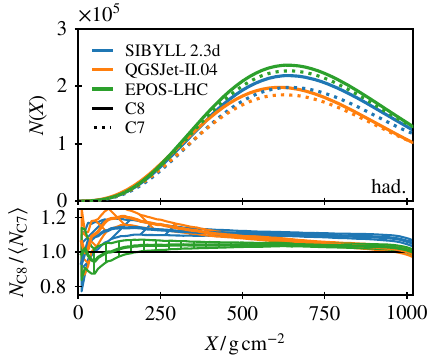}\put(71,74){\includegraphics[width=2.2cm]{watermark.pdf}}\end{overpic}
    \caption{Average longitudinal profiles of 300 air showers for various particle types (see indicator in plots) arising from hadronic cascades simulated with various high-energy hadronic interaction models and FLUKA as low-energy interaction model, both with CORSIKA~7 and CORSIKA~8. The hatched area shows the standard deviation of the mean.}
    \label{fig:longprofiles}
\end{figure}

\begin{figure}
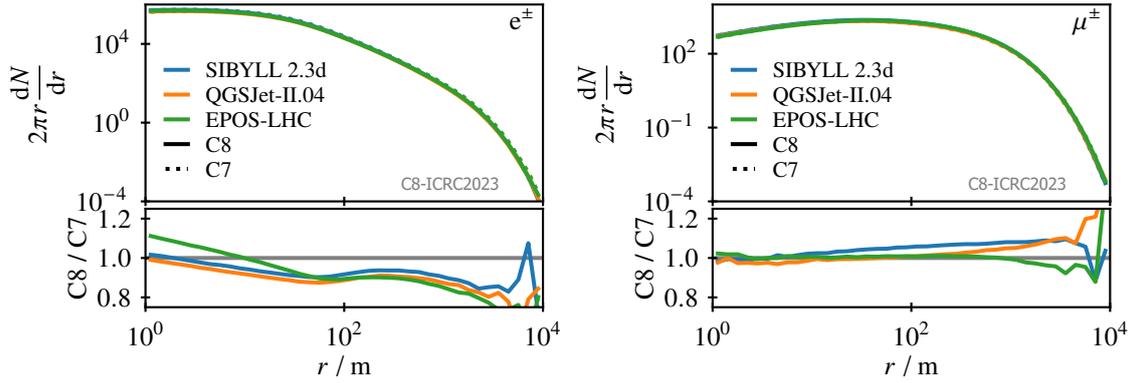

    \centering
    \begin{Overpic}{\input{ldf_e+-.pgf}}\put(65,30){\includegraphics[width=2.2cm]{watermark.pdf}}\end{Overpic}
    \begin{Overpic}{\input{ldf_mu+-.pgf}}\put(65,30){\includegraphics[width=2.2cm]{watermark.pdf}}\end{Overpic}
    \caption{Average lateral distributions of electrons and positrons (left) and muons (right) at ground for the same showers as shown in Fig.\ \ref{fig:longprofiles}.}
    \label{fig:ldfs}
\end{figure}

For the first time, we are now able to run full-fledged (ultra-)high energy hadron-induced air showers with thinning of the electromagnetic cascade in CORSIKA~8. In Fig.\ \ref{fig:longprofiles}, we showcase longitudinal distributions for the averages of 300 vertical proton-induced $10^{17}$\,eV air showers with thinning at the $10^{-6}$ level ($\wmax = \num{100}$, applied to all particles in CORSIKA~7; $\wmax = \num{50}$, only applied to the electromagnetic cascade in CORSIKA~8) with cuts for electromagnetic particles at \SI{10}{\MeV} and hadron/muon cuts at \SI{300}{\MeV}. The agreement between CORSIKA~8 and CORSIKA~7 for electrons/positrons, photons, muons and hadrons is within 10\%. We note a systematically higher number of muons and hadrons for CORSIKA~8 over CORSIKA~7, which also seems to be dependent on the high-energy interaction model. We will investigate the origin of these differences further in the future. We note that with Sibyll a larger spread between muon predictions across various codes has been observed previously~\cite{Reininghaus:2021box}. In Fig.\ \ref{fig:ldfs}, we show comparisons of the lateral distributions of electrons/positrons and muons at the ground. Again, we will investigate the observed differences more closely in the future.

\section{Radio-emission calculation}

From the start, radio-emission calculation has been a driver in the development of CORSIKA~8. This is, in particular, due to the radio-detection community's need for a flexible solution that can handle complex simulation scenarios, for example for in-ice radio detection experiments for which the radio emission from air showers crossing from air into ice is a very relevant background that can currently only be simulated by piecing together several simulation codes~\cite{UzairICRC2023}.

The implementation of the radio process in CORSIKA~8 has correspondingly been designed to decouple the emission calculation from the signal propagation through so-called \emph{Propagators} which handle all the complex transmission physics and multi-path propagation occurring in particular in dense media. This will allow easy incorporation of specific, complex use-cases required by the community. More details are provided in refs.~\cite{NikosICRC2023,AugustoICRC2023,JuanICRC2023}.

Radio emission, due to its coherent nature, is very sensitive to the exact energy and spatial distributions of electrons and positrons in an air shower. It thus also provides a very powerful diagnostic for the correctness of the electromagnetic cascade simulation. Since the ICRC2021, the results have improved tremendously; also, we are now able to simulate radio emission from air showers with CORSIKA~8 at high energies, thanks to the availability of thinning of the electromagnetic cascade, and with a fully realistic refractive index gradient of the atmosphere.

\begin{figure}
    \centering
    \begin{Overpic}[]{%
    \includegraphics[width=0.84\textwidth,trim={1.1cm 2.0cm 1.3cm 2.0cm},clip=true]{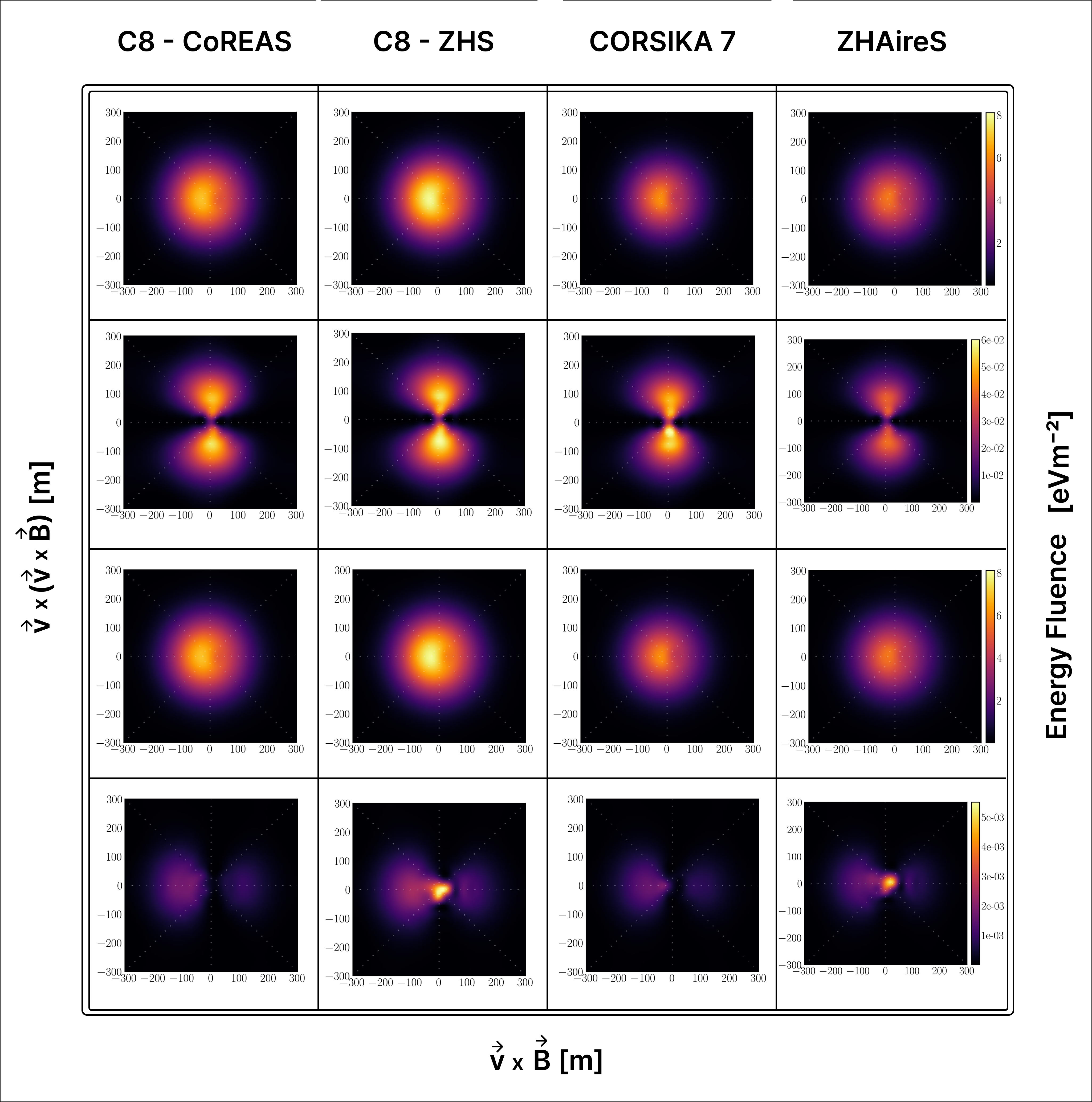}}
    \put(3,-1){\includegraphics[width=2.2cm]{watermark.pdf}}
    \end{Overpic}
    \caption{Fluence maps for 30-80\,MHz radio emission from a vertical $10^{17}$\,eV air shower simulated with both the endpoint (CoREAS) and ZHS formalisms in CORSIKA~8 are compared with simulations from CoREAS in CORSIKA~7 and with ZHAireS. From top to bottom, the rows show: total energy fluence, fluence in the $\vec{v} \times (\vec{v} \times \vec{B})$ (north-south), $\vec{v} \times \vec{B}$ (east-west) and $\vec{v}$ (vertical) polarizations. The $x$/$y$-axes show core distances along the $\vec{v} \times \vec{B}$ and $\vec{v} \times (\vec{v} \times \vec{B})$ directions. The color scales are identical within a given row.}
    \label{fig:fluencemap}
\end{figure}

CORSIKA~8 can perform concurrent simulation of the radio emission from the same particle cascade with two formalisms, the endpoint formalism as originally implemented in the CORSIKA~7 extension CoREAS~\cite{Huege:2013vt} and the ZHS formalism available in ZHAireS~\cite{AlvarezMunizCarvalhoZas2012}. Figure \ref{fig:fluencemap} shows energy fluence maps in the 30-80\,MHz band for observers at sea level for a $10^{17}$\,eV vertical iron-induced air shower in a horizontal \SI{50}{\micro\tesla} magnetic field, in a U.S.\ standard atmosphere with Gladstone-Dale refractivity gradient~\cite{GladstoneDale}, and with thinning at the $10^{-6}$ level with a maximum weight of 50 in CORSIKA~8 and settings leading to similar thinning in CORSIKA~7 and ZHAireS. The showers simulated with the different codes are very similar, but not fully identical; for details on the selected showers as well as the radio pulses and frequency spectra, we kindly refer the reader to~\cite{NikosICRC2023}. We note that the agreement between all four results (columns) in terms of symmetries is near-perfect. Earlier results~\cite{Karastathis:2021akf,NikosARENA2022} had shown significant deviations. Slight differences in the absolute strength still exist and will be investigated further. Another interesting finding is a ``blip'' of extra emission near the shower axis in the vertical signal polarization both seen in the ZHS-formalism simulation in CORSIKA~8 and the ZHAireS simulation, which is not seen in simulations with the endpoint formalism, however is not of practical relevance as it contributes at a very small absolute level.


\section{Conclusions and Outlook}

Since the ICRC2021~\cite{CORSIKA:2021czu}, we have made tremendous progress in the development of CORSIKA~8. Validation against CORSIKA~7 results, in particular, shows agreement generally on a $\approx$10\% level. Observed deviations will be investigated further in the future. We consider the code ``physics-complete'' and aim for a first expert-level beta release within the year of 2023. Especially for cross-media showers, urgently needed by the radio-detection community, CORSIKA~8 already now is the most flexible and complete solution available.

\let\oldbibliography\thebibliography
\renewcommand{\thebibliography}[1]{%
  \oldbibliography{#1}%
  \setlength{\itemsep}{1pt}%
}

{\footnotesize

\begin{thebibliography}{10}

\bibitem{HeckKnappCapdevielle1998}
D.~{Heck}, J.~{Knapp}, J.N.~{Capdevielle}, G.~{Schatz} and T.~{Thouw},
  \emph{{CORSIKA: A Monte Carlo Code to Simulate Extensive Air Showers}},  FZKA
  Report 6019, Forschungszentrum Karlsruhe (1998),
  \href{https://doi.org/10.5445/IR/270043064}{DOI}.

\bibitem{Engel:2018akg}
R.~Engel, D.~Heck, T.~Huege et~al., \emph{{Towards a Next Generation of
  CORSIKA: A Framework for the Simulation of Particle Cascades in Astroparticle
  Physics}}, \href{https://doi.org/10.1007/s41781-018-0013-0}{\emph{Comput.
  Softw. Big Sci.} {\bfseries 3} (2019) 2}
  [\href{https://arxiv.org/abs/1808.08226}{{\ttfamily 1808.08226}}].

\bibitem{CORSIKA:2021czu}
{\scshape CORSIKA} collaboration, \emph{{CORSIKA 8 -- Contributions to the 37th
  International Cosmic Ray Conference in Berlin Germany (ICRC 2021)}},  in
  \emph{{37th International Cosmic Ray Conference}}, 12, 2021
  [\href{https://arxiv.org/abs/2112.11761}{{\ttfamily 2112.11761}}].

\bibitem{JuanICRC2023}
J.~Ammerman-Yebra et~al., \emph{{Simulations of cross media showers with
  CORSIKA~8}}, \href{https://doi.org/10.22323/1.444.0442}{\emph{PoS} {\bfseries
  ICRC2023} (2023) 442}.

\bibitem{AlexanderICRC2023}
A.~Sandrock et~al., \emph{{Validation of Electromagnetic Showers in
  CORSIKA~8}}, \href{https://doi.org/10.22323/1.444.0393}{\emph{PoS} {\bfseries
  ICRC2023} (2023) 393}.

\bibitem{NikosICRC2023}
N.~Karastathis et~al., \emph{{Simulating radio emission from air showers with
  CORSIKA~8}}, \href{https://doi.org/10.22323/1.444.0425}{\emph{PoS} {\bfseries
  ICRC2023} (2023) 425}.

\bibitem{DominikICRC2023}
D.~Baack, J.-M.~Alameddine et~al., \emph{{Comparison and efficiency of GPU
  accelerated optical light propagation with CORSIKA8}},
  \href{https://doi.org/10.22323/1.444.0417}{\emph{PoS} {\bfseries ICRC2023}
  (2023) 417}.

\bibitem{Ferrari:2005zk}
A.~Ferrari, P.R.~Sala, A.~Fasso and J.~Ranft, \emph{{FLUKA: A multi-particle
  transport code (Program version 2005)}},  Tech. Rep. (2005),
  \href{https://doi.org/10.2172/877507}{DOI}.

\bibitem{Mucke:1999yb}
A.~Mücke, R.~Engel, J.P.~Rachen, R.J.~Protheroe and T.~Stanev, \emph{{SOPHIA:
  Monte Carlo simulations of photohadronic processes in astrophysics}},
  \href{https://doi.org/10.1016/S0010-4655(99)00446-4}{\emph{Comput. Phys.
  Commun.} {\bfseries 124} (2000) 290}
  [\href{https://arxiv.org/abs/astro-ph/9903478}{{\ttfamily
  astro-ph/9903478}}].

\bibitem{Engel:2019dsg}
F.~Riehn, R.~Engel, A.~Fedynitch, T.K.~Gaisser and T.~Stanev, \emph{{Hadronic
  interaction model Sibyll 2.3d and extensive air showers}},
  \href{https://doi.org/10.1103/PhysRevD.102.063002}{\emph{Phys. Rev. D}
  {\bfseries 102} (2020) 063002}
  [\href{https://arxiv.org/abs/1912.03300}{{\ttfamily 1912.03300}}].

\bibitem{AugustoICRC2023}
A.A.~Alves~Jr., N.~Karastathis, T.~Huege et~al., \emph{{Parallel processing of
  radio signals and detector arrays in CORSIKA~8}},
  \href{https://doi.org/10.22323/1.444.0469}{\emph{PoS} {\bfseries ICRC2023}
  (2023) 469}.

\bibitem{Sjostrand:2021dal}
T.~Sj\"ostrand and M.~Utheim, \emph{{Hadron interactions for arbitrary energies
  and species, with applications to cosmic rays}},
  \href{https://doi.org/10.1140/epjc/s10052-021-09953-5}{\emph{Eur. Phys. J. C}
  {\bfseries 82} (2022) 21} [\href{https://arxiv.org/abs/2108.03481}{{\ttfamily
  2108.03481}}].

\bibitem{Alameddine:2020zyd}
J.-M.~Alameddine, J.~Soedingrekso, A.~Sandrock, M.~Sackel and W.~Rhode,
  \emph{{PROPOSAL: A library to propagate leptons and high energy photons}},
  \href{https://doi.org/10.1088/1742-6596/1690/1/012021}{\emph{J. Phys. Conf.
  Ser.} {\bfseries 1690} (2020) 012021}.

\bibitem{Hillas:1981}
A.M.~Hillas, \emph{{Two interesting techniques for Monte-Carlo simulation of
  very high energy hadron cascades}},  in \emph{{Proc. 17th Int. Cosmic Ray
  Conf.}}, vol.~8, p.~193, 1981.

\bibitem{Kobal:2001jx}
M.~Kobal, \emph{{A thinning method using weight limitation for air-shower
  simulations}},
  \href{https://doi.org/10.1016/S0927-6505(00)00158-4}{\emph{Astropart. Phys.}
  {\bfseries 15} (2001) 259}.

\bibitem{Hansen:2010uk}
P.M.~Hansen, J.~Alvarez-Muñiz and R.A.~Vázquez, \emph{{A comprehensive study
  of shower to shower fluctuations}},
  \href{https://doi.org/10.1016/j.astropartphys.2010.11.001}{\emph{Astropart.
  Phys.} {\bfseries 34} (2011) 503}
  [\href{https://arxiv.org/abs/1004.3666}{{\ttfamily 1004.3666}}].

\bibitem{PhysRevD.83.014018}
S.~Ostapchenko, \emph{{Monte Carlo treatment of hadronic interactions in
  enhanced Pomeron scheme: {QGSJET-II} model}},
  \href{https://doi.org/10.1103/PhysRevD.83.014018}{\emph{Phys. Rev. D}
  {\bfseries 83} (2011) 014018}.

\bibitem{Pierog:2013ria}
T.~Pierog, I.~Karpenko, J.M.~Katzy, E.~Yatsenko and K.~Werner, \emph{{EPOS LHC:
  Test of collective hadronization with data measured at the CERN Large Hadron
  Collider}}, \href{https://doi.org/10.1103/PhysRevC.92.034906}{\emph{Phys.
  Rev. C} {\bfseries 92} (2015) 034906}
  [\href{https://arxiv.org/abs/1306.0121}{{\ttfamily 1306.0121}}].

\bibitem{Reininghaus:2023ctx}
M.~Reininghaus, T.~Sj\"ostrand and M.~Utheim, \emph{{Pythia~8 as hadronic
  interaction model in air shower simulations}},
  \href{https://doi.org/10.1051/epjconf/202328305010}{\emph{EPJ Web Conf.}
  {\bfseries 283} (2023) 05010}
  [\href{https://arxiv.org/abs/2303.02792}{{\ttfamily 2303.02792}}].

\bibitem{Reininghaus:2021box}
M.~Reininghaus, \emph{{The air shower simulation framework CORSIKA~8:
  Development and first applications to muon production}}, Ph.D. thesis, KIT \&
  UNSAM, 2022.
\newblock \href{https://doi.org/10.5445/IR/1000152097}{DOI}.

\bibitem{UzairICRC2023}
U.A.~Latif et~al., \emph{{Simulation of radio signals from cosmic-ray cascades
  in air and ice as observed by in-ice Askaryan radio detectors}},
  \href{https://doi.org/10.22323/1.444.0346}{\emph{PoS} {\bfseries ICRC2023}
  (2023) 346}.

\bibitem{Huege:2013vt}
T.~Huege, M.~Ludwig and C.W.~James, \emph{{Simulating radio emission from air
  showers with CoREAS}}, \href{https://doi.org/10.1063/1.4807534}{\emph{AIP
  Conf. Proc.} {\bfseries 1535} (2013) 128}
  [\href{https://arxiv.org/abs/1301.2132}{{\ttfamily 1301.2132}}].

\bibitem{AlvarezMunizCarvalhoZas2012}
J.~{Alvarez-Mu\~niz}, W.R.~{Carvalho Jr.} and E.~{Zas}, \emph{{Monte Carlo
  simulations of radio pulses in atmospheric showers using ZHAireS}},
  \href{https://doi.org/10.1016/j.astropartphys.2011.10.005}{\emph{Astropart.
  Phys.} {\bfseries 35} (2012) 325 }.

\bibitem{GladstoneDale}
J.H.~Gladstone and T.P.~Dale, \emph{{XIV. Researches on the refraction,
  dispersion, and sensitiveness of liquids}},
  \href{https://doi.org/10.1098/rstl.1863.0014}{\emph{Phil. Trans. Roy. Soc.
  Lond.} {\bfseries 153} (1863) 317}.

\bibitem{Karastathis:2021akf}
N.~Karastathis, R.~Prechelt, T.~Huege and J.~Ammerman-Yebra, \emph{{Simulations
  of radio emission from air showers with CORSIKA 8}},
  \href{https://doi.org/10.22323/1.395.0427}{\emph{PoS} {\bfseries ICRC2021}
  (2021) 427}.

\bibitem{NikosARENA2022}
N.~Karastathis, R.~Prechelt, J.~Ammerman-Yebra and T.~Huege, \emph{{Simulating
  radio emission from air showers with CORSIKA 8}},
  \href{https://doi.org/10.22323/1.424.0050}{\emph{PoS} {\bfseries ARENA2022}
  (2023) 050}.

\end{thebibliography}

\providecommand{\href}[2]{#2}\begingroup\raggedright\endgroup

}




\clearpage

\section*{The CORSIKA 8 Collaboration}
\small

\begin{sloppypar}\noindent
J.M.~Alameddine$^{1}$,
J.~Albrecht$^{1}$,
J.~Alvarez-Mu\~niz$^{2}$,
J.~Ammerman-Yebra$^{2}$,
L.~Arrabito$^{3}$,
J.~Augscheller$^{4}$,
A.A.~Alves Jr.$^{4}$,
D.~Baack$^{1}$,
K.~Bernl\"ohr$^{5}$,
M.~Bleicher$^{6}$,
A.~Coleman$^{7}$,
H.~Dembinski$^{1}$,
D.~Els\"asser$^{1}$,
R.~Engel$^{4}$,
A.~Ferrari$^{4}$,
C.~Gaudu$^{8}$,
C.~Glaser$^{7}$,
D.~Heck$^{4}$,
F.~Hu$^{9}$,
T.~Huege$^{4,10}$,
K.H.~Kampert$^{8}$,
N.~Karastathis$^{4}$,
U.A.~Latif$^{11}$,
H.~Mei$^{12}$,
L.~Nellen$^{13}$,
T.~Pierog$^{4}$,
R.~Prechelt$^{14}$,
M.~Reininghaus$^{15}$,
W.~Rhode$^{1}$,
F.~Riehn$^{16,2}$,
M.~Sackel$^{1}$,
P.~Sala$^{17}$,
P.~Sampathkumar$^{4}$,
A.~Sandrock$^{8}$,
J.~Soedingrekso$^{1}$,
R.~Ulrich$^{4}$,
D.~Xu$^{12}$,
E.~Zas$^{2}$

\end{sloppypar}

\begin{center}
\rule{0.1\columnwidth}{0.5pt}
\raisebox{-0.4ex}{\scriptsize$\bullet$}
\rule{0.1\columnwidth}{0.5pt}
\end{center}

\vspace{-1ex}
\footnotesize
\begin{description}[labelsep=0.2em,align=right,labelwidth=0.7em,labelindent=0em,leftmargin=2em,noitemsep]
\item[$^{1}$] Technische Universit\"at Dortmund (TU), Department of Physics, Dortmund, Germany
\item[$^{2}$] Universidade de Santiago de Compostela, Instituto Galego de F\'\i{}sica de Altas Enerx\'\i{}as (IGFAE), Santiago de Compostela, Spain
\item[$^{3}$] Laboratoire Univers et Particules de Montpellier, Universit\'e de Montpellier, Montpellier, France
\item[$^{4}$] Karlsruhe Institute of Technology (KIT), Institute for Astroparticle Physics (IAP), Karlsruhe, Germany
\item[$^{5}$] Max Planck Institute for Nuclear Physics (MPIK), Heidelberg, Germany
\item[$^{6}$] Goethe-Universit\"at Frankfurt am Main, Institut f\"ur Theoretische Physik, Frankfurt am Main, Germany
\item[$^{7}$] Uppsala University, Department of Physics and Astronomy, Uppsala, Sweden
\item[$^{8}$] Bergische Universit\"at Wuppertal, Department of Physics, Wuppertal, Germany
\item[$^{9}$] Peking University (PKU), School of Physics, Beijing, China
\item[$^{10}$] Vrije Universiteit Brussel, Astrophysical Institute, Brussels, Belgium
\item[$^{11}$] Vrije Universiteit Brussel, Dienst ELEM, Inter-University Institute for High Energies (IIHE), Brussels, Belgium
\item[$^{12}$] Tsung-Dao Lee Institute (TDLI), Shanghai Jiao Tong University, Shanghai, China
\item[$^{13}$] Universidad Nacional Aut\'onoma de M\'exico (UNAM), Instituto de Ciencias Nucleares, M\'exico, D.F., M\'exico
\item[$^{14}$] University of Hawai'i at Manoa, Department of Physics and Astronomy, Honolulu, USA
\item[$^{15}$] Karlsruhe Institute of Technology (KIT), Institute of Experimental Particle Physics (ETP), Karlsruhe, Germany
\item[$^{16}$] Laborat\'orio de Instrumenta\c{c}\~ao e F\'\i{}sica Experimental de Part\'\i{}culas (LIP), Lisboa, Portugal
\item[$^{17}$] Fluka collaboration
\end{description}

\vspace{-1ex}
\footnotesize
\section*{Acknowledgments}
This research was funded by the Deutsche Forschungsgemeinschaft (DFG, German Research Foundation) – Projektnummer 445154105.
Simulations were performed on the HPC cluster BinAC, supported by the High Performance and Cloud Computing Group at the Zentrum für Datenverarbeitung of the University of Tübingen, the state of Baden-Württemberg through bwHPC
and the German Research Foundation~(DFG) through grant no.\ INST~37/935-1~FUGG.

\end{document}